\documentclass[pra,onecolumn,superscriptaddress,nofootinbib]{revtex4}
\usepackage[a4paper, left=0.8in, right=0.8in, top=1.0in, bottom=1.0in]{geometry}

\usepackage[braket, qm]{qcircuit}
\usepackage[caption=false]{subfig}
\usepackage{pifont}
\usepackage{multirow}
\usepackage[utf8]{inputenc}
\usepackage[english]{babel}
\usepackage[T1]{fontenc}
\usepackage{amsmath}
\usepackage{amsfonts}
\usepackage{hyperref}
\allowdisplaybreaks

\usepackage{tikz}
\usepackage{braket}
\usepackage{natbib}
\usepackage{amsthm}
\usepackage{physics}

\usepackage[noend]{algpseudocode}
\usepackage{algorithm,algorithmicx}

\algrenewcommand\alglinenumber[1]{\sf\scriptsize\color{blue}{#1}}
\algrenewcommand\algorithmicrequire{\textbf{Input:}}
\algrenewcommand\algorithmicensure{\textbf{Output:}}
\begin{document}

\title{Patch-Based End-to-End Quantum Learning Network \\ for Reduction and Classification of Classical Data}

\date{\today}
\author{Jishnu Mahmud}
\email{jishnu.mahmud@bracu.ac.bd}
\affiliation{Department of Computer Science and Engineering, BRAC University, Dhaka, Bangladesh}
\author{Shaikh Anowarul Fattah}
\email{fattah@eee.buet.ac.bd}
\affiliation{Department of Electrical and Electronic Engineering, Bangladesh University of Engineering and Technology, Dhaka, Bangladesh}
\begin{abstract}
In the noisy intermediate scale quantum (NISQ) era, the control over the qubits is limited due to the errors caused by quantum decoherence, crosstalk, and imperfect calibration. Hence, it is necessary to reduce the size of the large-scale classical data, such as images, when they are to be processed by quantum networks. Conventionally input classical data are reduced in the classical domain using classical networks such as autoencoders and, subsequently, analyzed in the quantum domain. These conventional techniques involve training an enormous number of parameters, making them computationally costly. In this paper, a dynamic patch-based quantum domain data reduction network with a classical attention mechanism is proposed to avoid such data reductions, and subsequently coupled with a novel quantum classifier to perform classification tasks. The architecture processes the classical data sequentially in patches and reduces them using a quantum convolutional-inspired reduction network and further enriches them using a self-attention technique, which utilizes a classical mask derived from simple statistical operations on the native classical data, after measurement. The reduced representation is passed through a quantum classifier, which re-encodes it into quantum states, processes them through quantum ansatzes, and finally measures them to predict classes. This training process involves a joint optimization scheme that considers both the reduction and classifier networks, making the reduction operation dynamic. The proposed architecture has been extensively tested on the publicly available Fashion MNIST dataset, and it has excellent classification performance using significantly reduced training parameters.
\end{abstract}

\maketitle

\setcounter{secnumdepth}{0}
In this decade, which has seen a massive development of machine learning algorithms, convolutional neural networks (CNNs) have played a substantial role in image processing techniques. Many computer vision algorithms have been developed using CNNs and have had remarkable success in various applications, including image classification. Image classification is arguably the most fundamental problem of computer vision, having many uses in many disciplines, including but not limited to medicine, security, autonomous vehicles, and astronomy. As the computational power of modern-day computers increases, deeper neural networks with more parameters can be trained to capture the intricacies of the most complex data. Even with the increase in the computational abilities of modern-day computers, the training process for such large deep networks is proving computationally costly. Hence, the quest for developing alternative methods to implement computer vision tasks is deemed imperative.

Quantum computing has come into the limelight due to its remarkable computation abilities. Although most quantum computers of today are still noisy and susceptible to environmental impairments, the promise of exponential computational speedup and the development of near error-free quantum computing makes it a compelling competitor to be explored. Quantum versions of many machine learning algorithms have been proposed in recent times (\cite{cong_quantum_org}, \cite{Bausch2020RecurrentQN}, \cite{cherrat2022quantum}). One of the common advantages these algorithms share is their ability to capture the nuances of complex datasets by using a mere fraction of trainable parameters compared to their classical counterparts. The reduction of complexity by using a lower number of trainable parameters in optimizing the network is substantial.

Information from classical data, such as images, must be encoded into qubits before any quantum operation starts. Here, the classical data is mapped into quantum states in the Hilbert Space, which enables the network to perform quantum operations on the states, capturing more subtle patterns and intricacies in the dataset.  Superposition, one of the fundamental principles of quantum computers, comes into play in the first stage of mapping the classical data into qubits. Unlike a bit, which can only contain the values of 0 and 1, a qubit is a superposition of the two values that will collapse with probabilities into the eigenvectors of the operator that is being measured (\cite{yanofsky2008quantum}). Sequential quantum gates with trainable parameters are applied to the qubits in subsequent stages, exploiting quantum-unique phenomena, such as entanglement (\cite{yanofsky2008quantum}) and modulating their states. These states determine the probability of the qubits collapsing into a particular state after measurement.

The availability of near-term quantum computers marking the NISQ era fuelled much of the research in quantum machine learning using variational quantum circuits. These parameterized circuits are trained via classical learning algorithms, such as gradient descent, to update and achieve desired quantum states, which, in the following stages, predicted, classified, or extracted information from the classical data (\cite{farhi2018classification}, \cite{schuld2020circuit}, \cite{liu2018differentiable}). The main challenges of this particular genre of work are the choice of the architecture to use, selection of the parameterized quantum circuits (ansatzes) with which the architecture will be built, how to train its parameters, evaluation of the performance in terms of the expressibility of the quantum network, and achieving an accuracy/efficiency in carrying out the task. Various architectures such as \cite{cong_quantum_org}, \cite{perez2020data} have been proposed that are widely used for various classical data processing. Works such as those in \cite{maccormack2022branching}, \cite{opt-2qubit}, \cite{sim2019expressibility} provide insights into the entangling capability and expressibility of various configuration of ansatzes. 

Since the availability of quantum simulators and quantum computers to a broader range of researchers, optimization of this particular genre of works tailored for various applications can be observed in a wide array of fields such as medicine \cite{lung_cancer}, weather forecasting \cite{weather}, quantum chemistry \cite{chemical}. This particular sub-genre focuses on the modification and subtle tweaks in the quantum ansatzes and learning methods adopting various established architectures such as quantum support vector machines and quantum convolutional neural networks (QCNNs) or simply the addition of quantum pre/post-processing blocks incorporated in an already established classical algorithm (\cite{yang2022semiconductor}, \cite{yang2021decentralizing}).

Classification is a cornerstone of the problems tackled by quantum machine learning. Due to its relaxed demand for computational resources, it has been a prevalent problem tackled by quantum networks in this NISQ era, where the number of qubits and qubit interactions is limited. In this paper, a novel architecture is proposed with the aim of achieving three primary goals: 

\begin{itemize}
    \item developing a novel method for the classification of images using end-to-end quantum networks
    \item handling classical data of large size using NISQ quantum networks without any classical data reduction technique 
    \item addressing the problem of quantum decoherence experienced by circuits with increasing depth due to processing classical data of large size.
\end{itemize}

The dynamic quantum reduction network with classical self-attention is proposed to achieve these goals. The proposed network achieves excellent image classification results and challenges the variational networks in the current QML literature. The second aim stems from the imminent problem most QML research faces in this noisy quantum era. It can be observed that large data sizes such as images and longer speech samples are difficult to encode and process by a limited number of qubits unless a classical operation such as resizing or squeezing dimensions is performed, which supposedly loses some, if not a significant amount of information. Many recent works deploy classical autoencoders (\cite{ballard1987modular}), principal component analysis (PCA) (\cite{abdi2010principal}), or simple resizing as methods of reducing data sizes (\cite{tak_hur_boss}, \cite{fan2023hybrid}) before processing the data via various quantum networks. However, this method, in a way, takes aid from classical learning algorithms/preprocessing to preserve the information from higher dimensions, which then aids quantum learning processes. The use of autoencoders demands a high amount of classical processing with a large number of parameter training even before the quantum network comes into play. In the PCA technique, the spatial information of the image is lost alongside its significant computational complexities and decreased interpretability. Simple resizing or squeezing the image is much less demanding in terms of computational resources; however, naive down-sampling of image data loses valuable information. These methods raise questions regarding the contribution of the quantum network due to the addition of significant classical processing at the start.

Additionally, such classical techniques lose valuable local information as the reduction is performed on the whole image in one shot. Therefore, this paper proposes using a shallow quantum ansatz structure that reads through the small patches of the larger image, preserving local information to finally process them to construct a reduced version of the image that the classifier network can handle at low depth and small numbers of qubits. It is important to note two aspects of this reduction network. First, this network is trainable and learns the best way to summarize an image via joint optimization with the classifier. Second, a classical mask is created using statistical classical operations, which enriches the reduced quantum-processed data by providing a form of self-attention.

\begin{figure*}[t]
    \centering
    \includegraphics[width=0.95\textwidth]{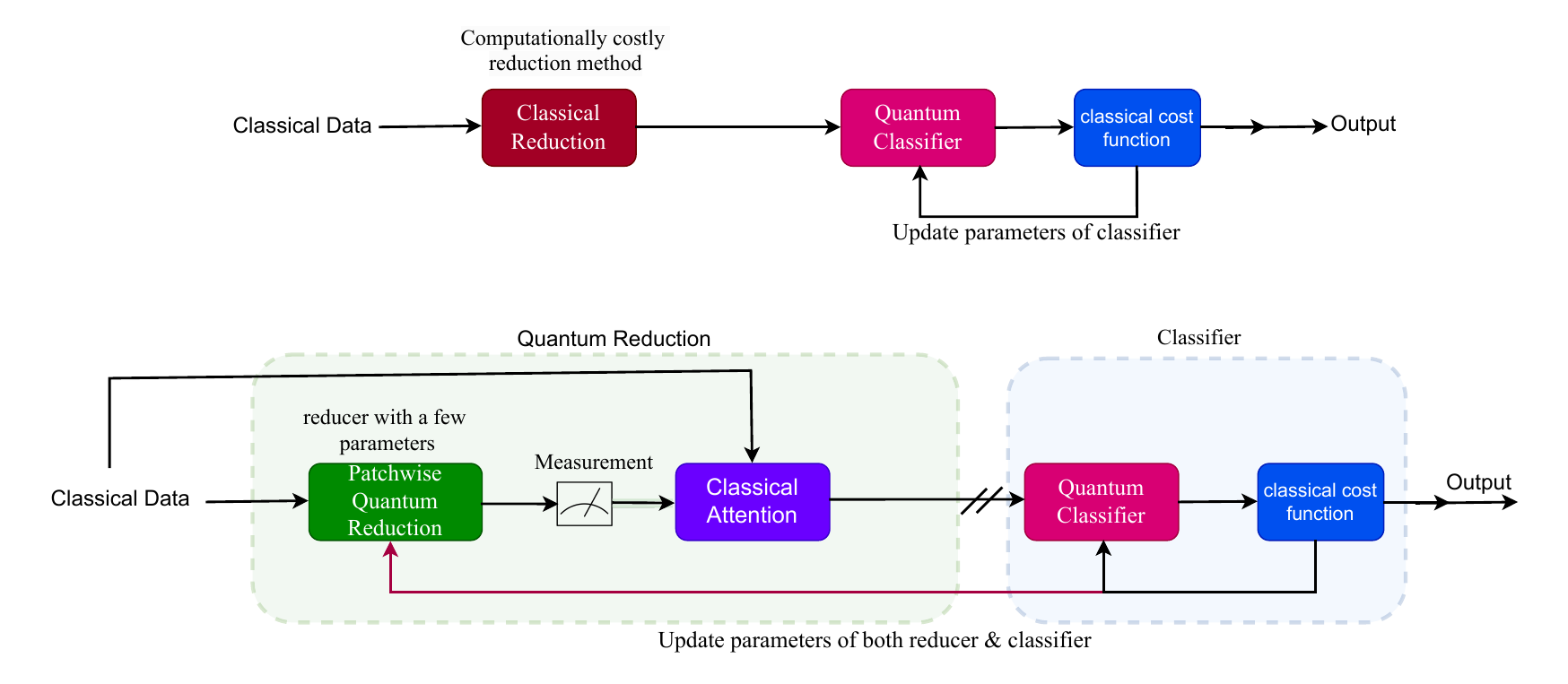}
    \caption{General simplified schematic for conventional QML networks (Top), versus the proposed network (Bottom). In conventional architecture, the classical reduction technique reduces the data per the quantum NISQ classifier, causing the quantum network to receive only selective classical information. Conversely, the quantum reducer is much less computationally demanding and is also jointly optimized with the learning process of the classifier. This method, therefore, decides which information to pass through to the classifier dynamically. The double slash between the quantum reduction and classifier block denotes that fact that all the patches in an image is processed sequentially and stored at that point before sent to the classifier.}
    \label{fig:intro_diagram}
\end{figure*}

Finally, the third aim is to arrest quantum decoherence as much as possible by reducing the depth of the network. This is particularly difficult considering that, as data size increases, the variational quantum circuit demands more qubits and simultaneously more depth to process those qubits before entering the classifying stage. In this NISQ era, much of the QML research justifies using classical means to reduce data sizes so that the quantum network can assert an upper bound to the number of qubits and, therefore, circuit depth to tackle quantum decoherence. However, since this paper focuses on using quantum ansatzes only to reduce and process quantum states, the network's structure must be such that both the quantum reduction network and the classifier block are separated by qubit measurement and eventually jointly optimized using a joint objective function. This unique design makes quantum networks shallow and more flexible when selecting different network architectures as the final classifiers. The comparison between the conventional QML method and the proposed architecture is illustrated in figure \ref{fig:intro_diagram} using simplified block diagrams.

Since the goals of the proposed network are clear, it is now essential to provide a bird’s eye view of its physical structure. This work proposes a reducer-classifier structure based on a new quantum image reduction method coupled with a quantum classifier. The network's novel architecture, illustrated using a block diagram in figure \ref{fig:overall_diagram}, combines the essence of QCNNs with the reducer-classifier structure. This reducer structure, which preserves the local information of the image by processing it via small patches, is shown to outperform the conventional classical reduction techniques, such as a classically trained autoencoder or simple resizing (downsampling). The reducer-classifier structure has quantum attributes in all the layers and only relies on the classical cross-entropy cost function for gradient descent to update the trainable parameters of the reducer and the classifier. A joint optimization is performed on the quantum reducer and classifier, minimizing the classical cost between the labels and the predicted values. It can also be observed that after the parameters of the reduction blocks have been trained, the reducer part of the trained model can be used to create better feature representations of the classical data, such as images in the reduced sizes. 

The proposed architecture should not only be interpreted as a network for classifying classical images but also as an effective size-reduction technique. This technique is of immense importance, especially in the NISQ era, for handling data of considerable size using the available quantum networks. The proposed architecture further addresses the issue of increased quantum decoherence experienced by circuits when processing large classical data due to the increased depth of the quantum network. An extremely shallow reducer is designed to process the image patches, which attempts to keep decoherence to a minimum.

The rest of the paper is arranged as follows: The following section provides an in-depth discussion of the overall network architecture proposed in this paper. This section is further divided into smaller sub-sections, where the reduction system, the classifier system, and finally, the join-optimization process are highlighted. Next, the paper includes the results section, where detailed discussions on the dataset, the simulation environment, and the simulation results are included. The model is further tested in the simulation result subsection with variations in the classifier, and reducer architecture to test the proposed architecture. Finally, concluding remarks are made regarding this work in the last section, and scopes for future work are discussed.

\section{Architecture}
The proposed architecture consists of a quantum reduction block that receives the classical data and performs a quantum operation patch-wise to reduce the image into a quantum-processed matrix. The quantum reduction network preserves the local information by processing each patch using a sequence of quantum convolutional and pooling layers to summarize the information to a single quantum-processed representative value. Furthermore, a classical attention mechanism is used to incorporate the information obtained by the classical reduction of an image. In particular, a local normalized average pooling is performed on the image, which creates a self-attention mask that enriches the quantum-reduced matrix further. As a result, the proposed reduction technique utilizes the information of both quantum and classical reduction while significantly reducing the computational costs of conventional reduction methods. Moreover, it must be noted that the proposed quantum reduction system has trainable parameters that are jointly optimized with the quantum classifier, making the proposed reduction dynamic with respect to the final objective function.



\begin{figure*}[t]
    \centering
    \includegraphics[width=0.85\textwidth]{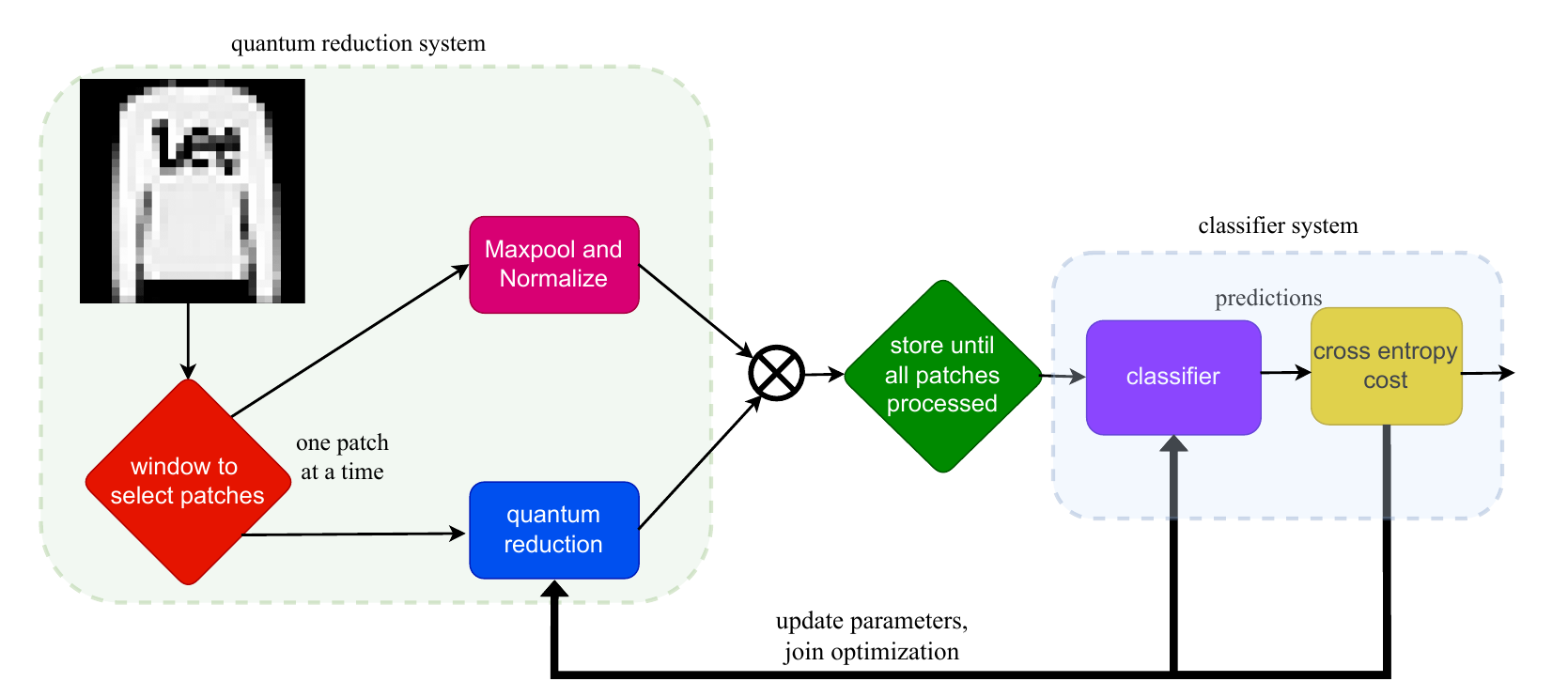}
    \caption{Simplified block diagram of the proposed architecture. The first block patches the images, which selects a section of size $r \times r$. The classical processing (Maxpool and Normalize) and the quantum reducer block process each patch in parallel. The values derived from the parallel lines are multiplied and stored until all the patches are processed. Hence, a feature matrix is constructed and processed by the quantum classifier to make predictions post-measurement. The predicted vectors are, consequently, used by the classical cross entropy function to generate updated parameters for both the quantum reducer and classifier block to update the parameters to minimize cost.}
    \label{fig:overall_diagram}
\end{figure*}

The classifier waits till the reducer has processed all the zones of the image, collects this information, and processes it through a series of variational quantum gates to separate the classes. Finally, upon the measurement of the qubits, the network predicts the class of each image and optimizes a classical cost function, which modulates the free parameters in both the reducer and the classifier network, thus performing a joint optimization of both quantum networks from the cost incurred from the prediction and ground truth labels. A detailed analysis of each subsystem engaged in this novel architecture is done in the following sub-sections. A simplified block diagram of the whole architecture is shown in figure \ref{fig:overall_diagram}. 


\subsection{The dynamic quantum reduction system}
Initially, the reducer system comprises a window that slides through the image, capturing pixel values (patches) and forwarding those as input to the reducer network. The quantum reduction network passes these values through amplitude encoding, which maps them to the quantum states in the Hilbert Space. Consequently, the states pass through the sequence of ansatzes with trainable parameters, where the number of qubits is decreased sequentially using quantum pooling layers until one final qubit is measured. Therefore, the patch the window captures is reduced to a representative single value, which is expected to summarize the information in the image patch better as the joint training continues. 
The major challenge in designing this network is selecting ansatzes that can learn the intricate information (readily trainable) from the patch and represent it in a single value, preserving the most important nuances in the data needed for classification.

\subsubsection{The Window}

As the preliminary step, a dataset containing images of size $k\times k$ is padded to increase the size to $N\times N$, where $N^2$ is a power of 2. Indeed, this step is not necessary $k^2$ is already a power of 2. This step is essential as amplitude encoding the classical pixel values in amplitude encoding require them to be a power of 2. 

The reducer selects a patch of pixels of size $r\times r$ $(r<N)$ in a row-priority sequential manner with a stride of r, meaning there is no overlapping of pixels on two adjacent windows. Consequently, the image is fragmented into $2^n = (\frac{N}{r})^2$ number of $r\times r$ $(= 2^m)$ pieces, each taken as input by the reducer system.

\begin{figure*}[t]
    \centering
    \includegraphics[width=0.95\textwidth]{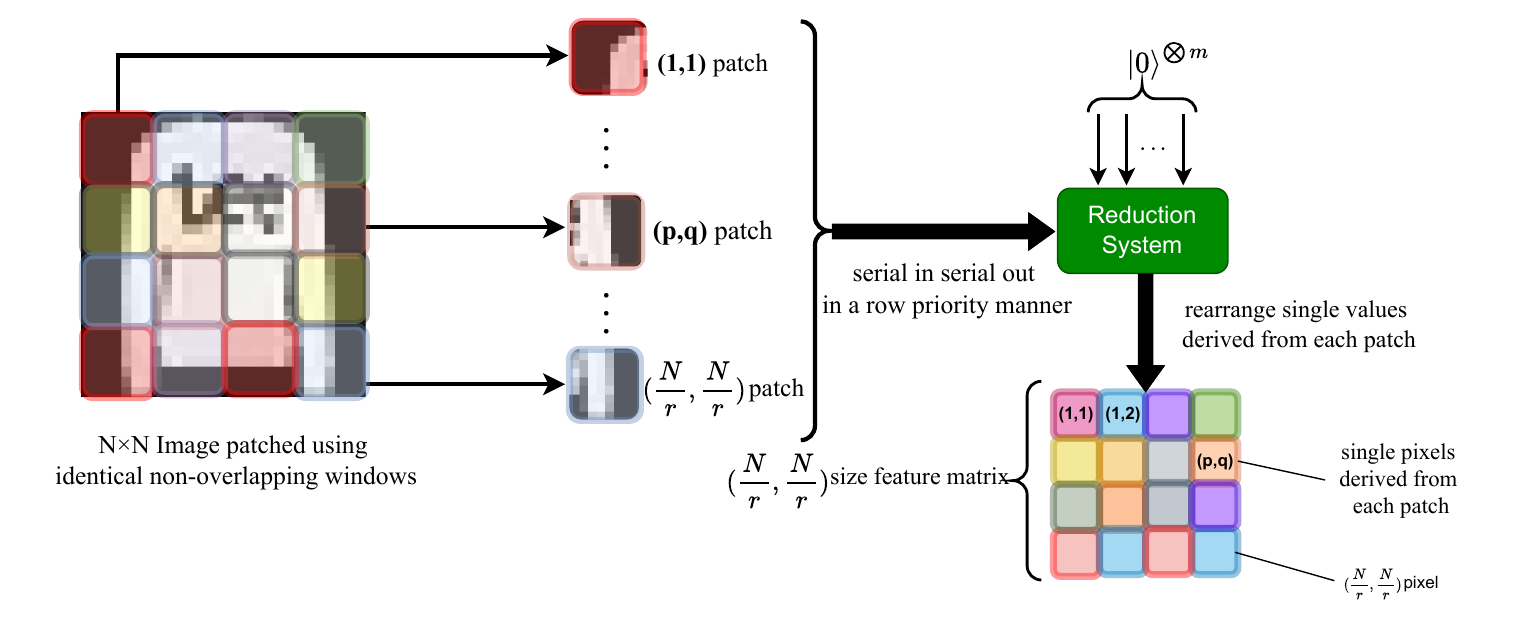}
    \caption{An illustration for how $r\times r$ sized patches are made from the image and consequently sent to the reducer serially to be reduced into a reduced feature matrix. The (p,q) patch from the original is processed to produce the (p,q) element in the reduced feature matrix.}
    \label{fig:sliding_window}
\end{figure*}

The quantum reducer system, now, is responsible for processing the window of $r\times r$ pixels via the application of a series of ansatzes and sequential reduction of qubits to represent the patch by a single value. This procedure is graphically represented in figure \ref{fig:sliding_window}. As illustrated, each patch is labeled with respect to its Cartesian coordinate position in the original image, and it can deduced that the reduction network processes the $(p,q)$ patch to produce the $(p,q)$ pixel element in the final feature matrix. Indeed, the values extracted from the patches are reshaped to form this feature matrix, where each final value occupies the coordinate of its parent patch.

The reduction network can be further divided into the quantum and classical processing subsystems; the initial part of the quantum reduction network consists of an amplitude encoding block, which is responsible for encoding the qubits with the fragmented classical data from each of the windows. In this step, the process of projecting the features of the input data received from the previous layer into quantum states, commonly known as quantum feature encoding, is performed. Mathematically, the mapping of the classical input data denoted as $X$, is mapped to a quantum state residing in the Hilbert Space. Amplitude embedding is one of the many ways classical data can be mapped into quantum states, and it does so by encoding $2^m$ classical data points into $m$ qubits. In this case, the number of data points is $r^2$ (directly from $r\times r$), and therefore, $m = \log_2{(r^2)}$ qubits are used in the reducer network, which can be observed as being input to the encoding block in figure \ref{fig:sliding_window}. The qubits are passed through a series of quantum convolutional and pooling layers inside the reduction circuit.

In quantum image processing literature, the image is expected to be flattened, and consequently, the resulting one-dimensional vector is to be processed directly through amplitude encoding, resulting in the whole image being input into the network via a single pass. However, this method fails to preserve the spatial nature of images. It can be argued that the images' spatial information is unimportant if the network can achieve competitive results via a direct pass. Even in that case, the dimension and size of the image (or any other classical data) will pose a significant obstacle as there is a theoretical upper limit to the amount of data that can be encoded in a finite-dimensional quantum state (\cite{superdense}). This shows that the number of qubits needed will increase with the size or number of the classical data points, which is undesirable in this NISQ era. This problem is addressed and avoided in the proposed architecture by processing individual patches sequentially to preserve spatial features and, therefore, developing a method to accommodate large classical data in quantum networks, which is imperative in image-processing tasks.

\subsubsection{The quantum reduction network}\label{subsec2}

\begin{figure*}[t]
    \centering
    \includegraphics[width=0.85\textwidth]{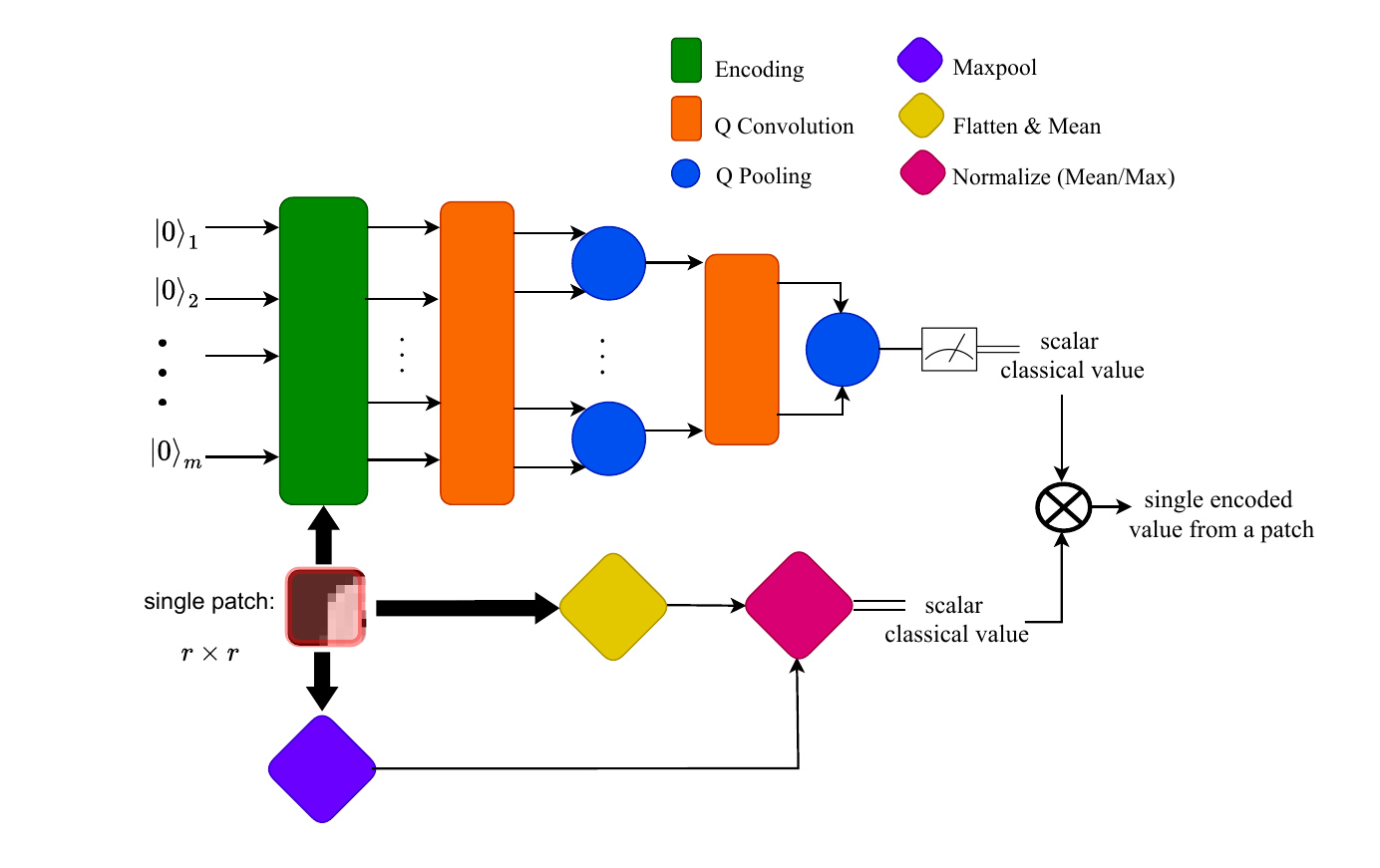}
    \caption{Simplified schematic of the complete reducer network. After the sliding window makes the patches, they will be processed by the classical and quantum reduction blocks. The quantum reduction block has sequential quantum convolutional and pooling layers in the conventional tree-like structure, reducing the dimension to a single scalar via the measurement of a single qubit reduced from $m$ qubits. The classical block processes the patch using the mean value and divides it by the local maximum to produce a scalar representation.}
    \label{fig:encoder_network}
\end{figure*}

The overall quantum reduction network is shown in figure \ref{fig:encoder_network}. The classical pixel values from each window flow into the m-qubit reduction network, where they are initially passed through amplitude encoding. The amplitude encoding block maps the classical values of the pixel to quantum states by
\begin{equation}
    \phi : X\mapsto H\
\end{equation}
where $X$ represents the values in the classical domain, and $H$ represents the Hilbert Space, which encompasses the quantum states. As the name suggests, amplitude encoding modulates the amplitudes of the $2^m$ possible basis states created by the $m$ qubits. 
The following equation shows that for $m$ qubits, which exist as a superposition of their basis states (trivially $\ket{0}$ and $\ket{1}$), there will be $2^m$ terms. Each will have an amplitude value that $2^m$ classical pixel values can modulate \cite{schuld2018supervised}. The composite state of $m$ qubits out of the amplitude encoding block can be expressed by the following equation
\begin{equation}
   \ket{\psi_{qubits}} = \sum_{a_{m},.. a_{2}, a_{1} \in \{0,1\}} C_{a_{m}...a_{1}}\ket{a_{m}a_{m-1}...a_{2}a_{1}}
\end{equation}
where $a_{m}$, $a_{m-1}$... $a_{2}$, $a_{1}$ represent the basis state of the $4$ qubits, $C_{a_{m}...a_{1}}$ represent the amplitude modulated by the pixel values, and $\ket{\psi_{qubits}}$ represent the composite state of the four qubits after amplitude encoding.

\begin{figure}[t]


    \centering
    \[ \Qcircuit @C=1.2em @R=2.1em {
    \lstick{\ket{\psi_1'}} & \qw & \gate{R_z(\theta_1)} & \qw & \qw & \qw & \gate{R_x(\theta_2)} & \rstick{\ket{\psi_1''}} \qw\\
     \lstick{\ket{\psi_n'}} & \qw & \ctrl{-1} & \qw & \gate{X} & \qw & \ctrlo{-1} & \qw \qw
    } \]
    \caption{The ansatz is used to build the pooling layer in the reduction block and the classifier.}
    \label{fig:ansatz}
\end{figure}
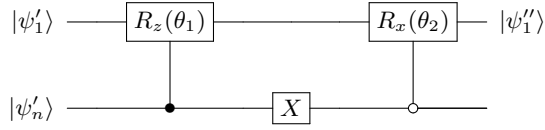

Each quantum convolutional layer is constructed using parameterized gates and cascaded to each other, as visualized in figure \ref{fig:conv_and_pooling_layers}. These parameters of the quantum gates are to be updated using gradient descent once the cost function is calculated using the predictions and labeled data. The ansatz or parameterized gates used to construct the convolutional and pooling layers are illustrated in figures \ref{fig:encoding_conv_ansatz} and \ref{fig:ansatz}. The quantum pooling layer, just like its classical counterpart, reduces the data, and it does so by taking two qubits as input and eliminating one of them after the gate operations shown in figure \ref{fig:conv_and_pooling_layers}. The following equation can describe the change in state in any given layer:

\begin{figure*}[t]
    \centering
    \includegraphics[width=0.85\textwidth]{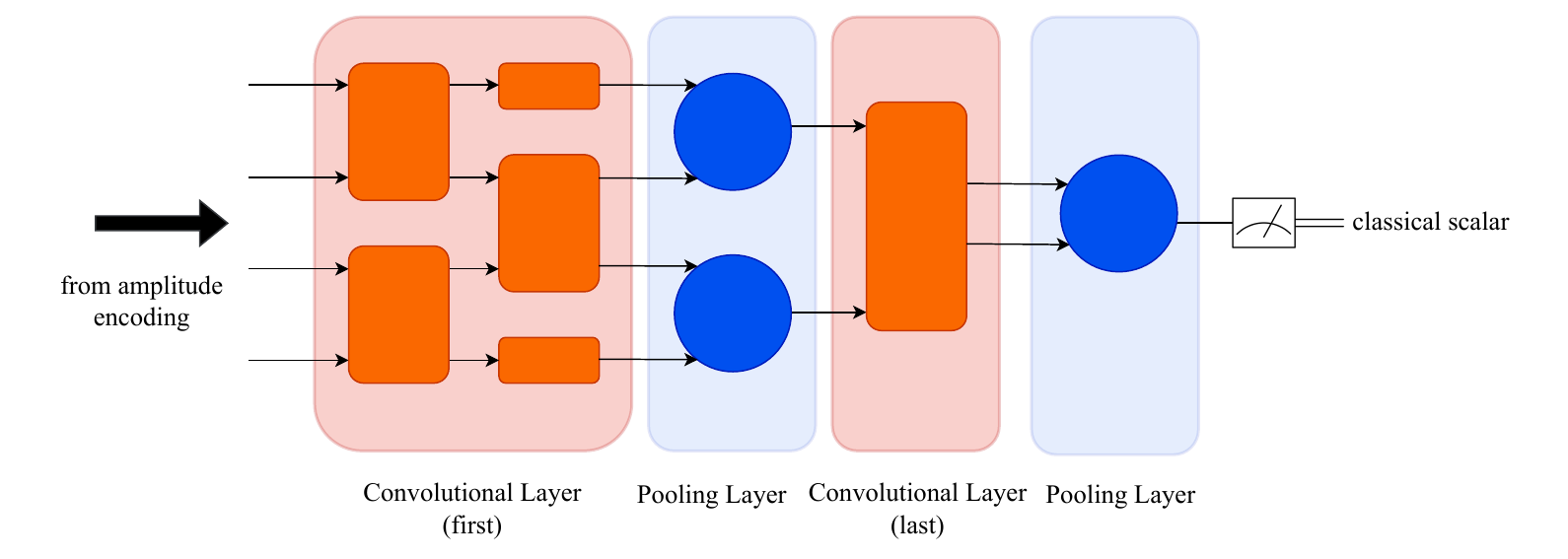}
    \caption{The internal structure of the convolutional and the pooling layers inside the quantum reducer block is shown. A sequence of quantum convolutional and pooling layers reduces the number of qubits. The quantum convolutional layers are connected using ansatzes, as shown in the quantum first convolutional layer block. The quantum convolutional layer is connected only when the number of qubits is reduced to $2$, as shown in the convolutional layer's last block. The pooling layers use the pooling ansatz that takes input $2$ qubits and outputs $1$ qubit. The convolutional and pooling layers comprise the ansatzes explicitly shown in figure \ref{fig:encoding_conv_ansatz} and \ref{fig:ansatz}.}
    \label{fig:conv_and_pooling_layers}
\end{figure*}


\begin{equation} 
  \rho_{i+1} =  \mathrm{Tr}_{A_i} \left( U_{\theta_{i}} \rho_{i} U_{\theta_{i}}^{\dagger}\right)
\end{equation}
where $\rho_{i}$ is the density matrix representation of the input state, $\rho_{i+1}$ is the density matrix representation of the output state of the layer, $U_(\theta)$ is the parameterized unitary operation of the layer, and $\mathrm{Tr}_{A_i}(\cdot)$ is the partial trace operation over subsystem $A_i$. This derives the reduced state of the system, excluding any desired subsystem denoted by  $A_i$. It must be noted that $\big|\psi\rangle$ is the composite space of $n$ qubits involved in the system, $\big|\psi\rangle^{\otimes n}$. In simpler terms, using state-vector formalism, the output state $\ket{\psi_{conv_{out}}}$ from $\ket{\psi_{conv_{in}}}$ passing through the quantum convolutional block can be expressed in the following way:
\begin{equation}
    \ket{\psi_{{conv}_{out}}} = U_{conv} \ket{\psi_{{conv}_{in}}}
\end{equation}
The $U_{conv}$ is the resultant unitary operator of the convolutional layer. This equivalent operator, $U_{conv}$, can be derived from the individual operators, a direct consequence of the ansatz used for the convolutional layers. One of the ansatz used for the construction of the quantum convolutional layers on the reducer side (Fig. \ref{fig:encoding_conv_ansatz}) is given by 
\begin{equation}
    U_{1,2} =  (R_{x2}(\theta) \otimes R_{x1}(\theta)) \star Cx_{2,1} \star (H_{2} \otimes H_{1}) \label{encoder_ansatz}
\end{equation}
where $H_{i}$ represents the Hadamard operation on the i\textsuperscript{th} qubit, $Cx_{2,1}$ is the controlled not (Cnot) operation performed on qubit 2 as target and 1 as control, and $R_{xi}$ represents the rotation operation in the $x$ direction on the ith qubit. This ansatz is inspired by its use in parameterized quantum circuits in tree tensor networks \cite{grant2018hierarchical}. Furthermore, the parameterized ansatz contains only two training parameters, which keeps the computational cost of the reduction circuit to a minimum and maintains a very shallow network depth. This aspect is particularly advantageous if the trained reducer is used to reduce the size of images or other classical data to enriched representative matrices, utilizing low depth and computational cost, which can consequently be used more efficiently by various classical and quantum post-processing networks for many applications.


\begin{figure}[t]
    \centering
    \[\Qcircuit @C=1em @R=1.8em {
        & \gate{H} & \ctrl{1} & \gate{R_x(\theta_1)} & \qw \\
        & \gate{H} & \targ & \gate{R_x(\theta_2)} & \qw \\
    }\]
    \caption{The ansatz used for constructing the convolutional layers in the reduction/classifier subsystem. It contains only 2 trainable parameters.}
    \label{fig:encoding_conv_ansatz}
\end{figure}
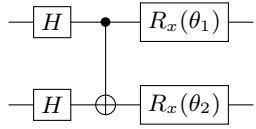

If the reducer network processes a patch of size $r \times r$, amplitude encoding can encode the classical information in $m$ $=(\log_{2}(r^2))$ qubits. Hence, the equivalent convolutional operator $U_{conv}$ of the convolutional layers can be derived directly from the equation \ref{encoder_ansatz}. Therefore, for the first convolutional layer, the input $\psi_{{conv}_{in}}$ transforms into $\psi_{{conv}_{out}}$ via the unitary operator
\begin{equation}
    U_{conv} = (U_{1,m} \otimes U_{2,m-1} \otimes ... \otimes U_{\frac{m}{2},\frac{m+2}{2}})\star(U_{1,2} \otimes U_{3,4} \otimes ... \otimes U_{m-1,m})
\end{equation}
where $U_{p,q}$ refers to the unitary in equation \ref{encoder_ansatz}.
It must be clarified that the $U$ is the operation of the ansatz on two qubits, and the $U_{conv}$ is the equivalent operation of the convolutional layer on $m$ qubits (as denoted by the tensor product). The $m$ qubits are sequentially reduced to a single qubit by the pooling layers after each quantum convolutional layer. These operations can be mathematically expressed by the following equations.
\noindent The pooling ansatz, illustrated in figure \ref{fig:ansatz} is denoted by
\begin{equation}
    P_{1,2} = Cz_1(\theta_1) \star I \otimes X_2  \star  Cx_1(\theta_2)
\end{equation}
where $Cz_1$ refers to the controlled $Z$ operation on qubit 1, $X_2$ is the $X$ operation on qubit 2 and $Cx_1$ is the controlled $X$ operation on qubit 1. Therefore, the following denotes the pooling layer on the $m$ initial qubits.
\begin{equation}
    U_p = \bigotimes_{i=1}^{m} P_{2i-1,2i} 
\end{equation}
The elimination of the qubits is expressed as
\begin{equation}
      \rho_{i+1} =  \mathrm{Tr}_{A_i} \left( U_{p} \rho_{i} U_{p}^{\dagger}\right)
      \label{pooling_trace}
\end{equation}
where $\mathrm{Tr}_{A_i}$ refers to the trace over subsystem $A_i$ which is the first qubit from each two-qubit pooling ansatz in the pooling layer, and $\rho_{i}$ and $\rho_{i+1}$ denotes the density operators of the states input to and output from the pooling layer respectively. Finally, the expectation value on the Pauli-Z operator is measured. The measurement of the final qubit is denoted mathematically as 
\begin{equation}
    y_{measured} = \bra{\psi_{out}}Z\ket{\psi_{out}}
    \label{expectation_value}
\end{equation}
This outputs the expectation value of the qubit on the Pauli-Z observable. The Pauli-Z is chosen because it has two eigenvalues of $1$ and $-1$ corresponding to the eigenvectors $\ket{0}$ and $\ket{1}$. Therefore, the quantum state of the final qubit will collapse into either eigenvector, and the measurement value will be the corresponding eigenvalue. When the same quantum network is run for all the window values spanning the whole image, it results in $(\frac{N}{r})^2 $ values, each derived from a unique $r\times r$ image segment. Therefore, to summarize, the image of size $N\times N$ is converted to a vector of length $(\frac{N}{r})^2$, then reshaped to a quantum feature matrix of size $\frac{N}{r}\times \frac{N}{r}$. Each value takes the position of its original $r\times r$ origin matrix, creating a quantum feature extracted matrix with decreased size and a degree of spatial feature preservation.

\subsubsection{The self-attention mechanism}
When summarizing the necessary information in the image, it is essential that some classically derived information from each patch is considered in the reduction block. The use of statistical data, such as pooling and normalization, adds self-attention to each quantum-derived patch value, serving as a reminder of the original classical local information in the raw image. At the same time, this method cleverly bypasses any training of classical networks and avoids significant computational costs.

With this aim in mind, in parallel to the quantum reduction operation in the previous section, the ($r\times r$) patches from the input image are sent to the max pool and normalize block where classical average pooling is performed on the patch, and the global maximum value of the pixel of the given patch normalizes the pooled values. Thus, for the $N\times N$ image, there will be $(\frac{N}{r})^2 $ - $r\times r$ patches with $r^2$ pixel values in each of them. Each $r\times r$ patch, $X_{p,q}$ is uniquely identified by a $(p,q)$ (Cartesian) coordinate, which is its position on the reduced classical feature matrix. Initially, the patches are flattened into a $r^2\times1$ vector, which then undergoes the following classical operation:
\begin{equation}
    Y_{p,q} = \sum_{i} \frac{X_{p,q,i}}{max_{i}(X_{p,q})\sum_{i}i}
\end{equation}
where $Y_{p,q}$ is a scalar value for the $(p,q)$ patch of the reduced classical feature matrix. It is derived from $X_{p,q, i}$, which is the i\textsuperscript{th} element of the patch after flattening, and $max_{i}(X)$ is the maximum among those elements. Once this procedure has been done for all the patches, the classical attention mask is created by reshaping the output, i.e., the $Y_{p,q}$ values, into an $\frac{N}{r}\times \frac{N}{r}$ matrix which forms a reduced representation of the original image. A single value from each of the $r \times r$ matrices means that this method reduces a single image of size $N\times N$ to $\frac{N}{r}\times \frac{N}{r}$, which is exactly the size of the resulting image from the quantum reduction block. 
The quantum and the classical feature matrix are then multiplied to create what is proven to be a much better representation of the larger image in reduced form. The following relation can now express each element of the matrix:
\begin{equation}
    \Gamma_{p,q_{p,q\in [1-r]}} = \bra{\psi_{out-p,q}}Z\ket{\psi_{out-p,q}} 
    \times \sum_{i} \frac{X_{p,q,i}}{max_{i}(X_{p,q})\sum_{i}i}
\end{equation}
Where $\Gamma_{p,q}$ is the classical value derived from the sequential classical-quantum processing and $\psi_{out-p,q}$ is the quantum state out of the reduction subsystem processing the $(p,q)$ patch. $Z$ is the Pauli-Z observable, $X_{p,q}$ is the flattened $(p,q)$ patch, and $X_{p,q, i}$ is its i\textsuperscript{th} element of the patch.

This synthesized quantum-classical reduced matrix preserves spatial information, which is essential in image processing while utilizing the various advantages of quantum feature extraction in its ability to extract nuanced information from classical data. The problems caused by quantum decoherence become more imminent with increasing circuit depth, which is demanded when processing large classical data. The proposed quantum reduction method recognizes this lingering issue and uses shallow networks to process image patches instead of trying to encode the whole image at once. This enables the network to be extremely shallow and thus aids in keeping quantum decoherence problems to a minimum. Combining classical operations and a shallow parameterized quantum reducer structure to derive an enriched but reduced representative matrix from classical images is a first, to the author's knowledge.


\subsection{The classifier system}

\begin{figure*}[h]
    \centering
    \includegraphics[width=1\textwidth]{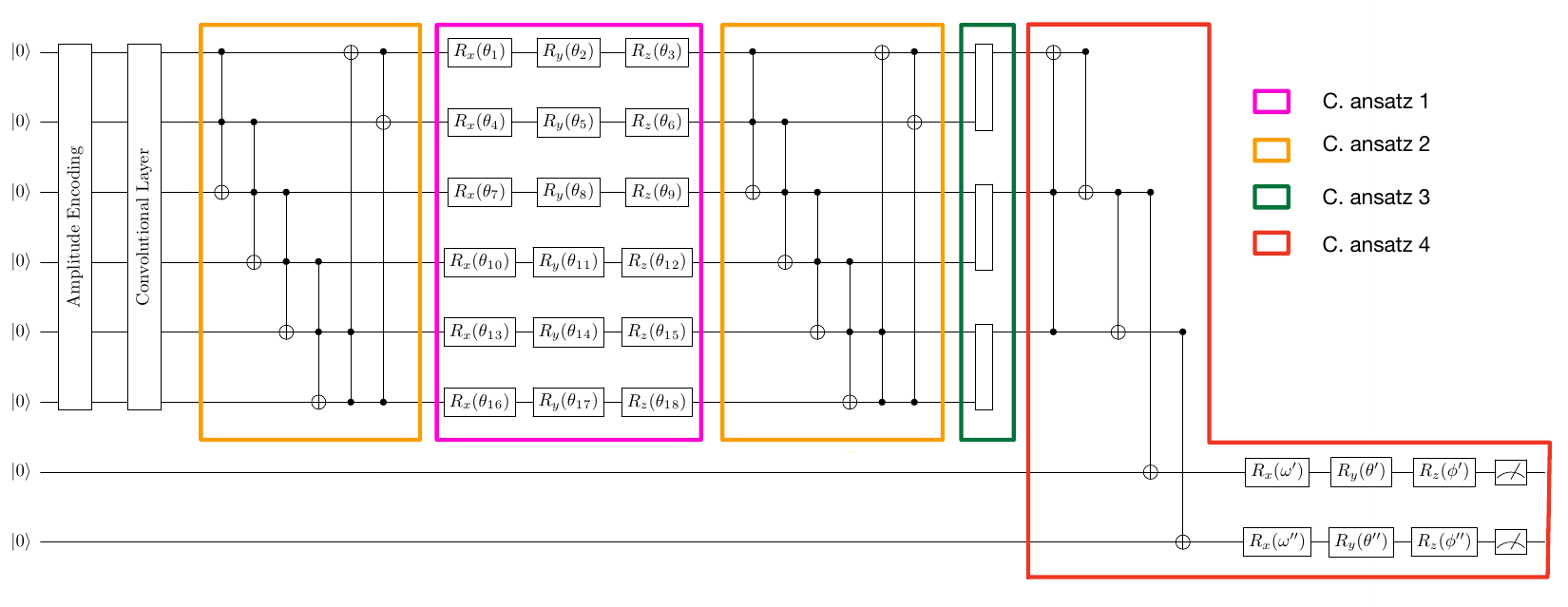}
    \caption{The overall diagram used as the quantum classifier processes the reduced feature matrix. The $n=6$ qubits encode the $2^n=64$ size classical feature matrix into a $2^n$ dimensional quantum state. The state is passed through several quantum ansatzes into the final step, where the qubits interact with the auxiliary qubits, which are finally measured. The number of auxiliary qubits needed depends on the number of classes used to classify using the network. The trainable parameters are shown using $\theta$ in the circuit. It must be noted that the convolutional layer also has trainable parameters and possesses the same structure as the convolutional layer of the reduction network illustrated in Fig \ref{fig:conv_and_pooling_layers}}
    \label{fig:decoder_network}
\end{figure*}

The remarkable aspect of the proposed quantum reduction block, since it produces a reduced classical feature matrix, is that it reserves the flexibility to choose any classical or quantum classifier for joint optimization, providing flexibility in application. Although, primarily, a quantum classifier structure is proposed to receive the feature matrix and process it into decision labels of a particular class, a classical neural network structure (for example, a sequence of linear layers) or a completely different quantum classifier network can be used to perform a join-optimization of the reducer and the classifier parameters. Once the proposed quantum reduction network has been trained (with any quantum/classical classifier), it can be subsequently used to produce feature-rich matrices at a reduced size from the original images.

The reduced feature matrix is taken as input in the quantum classifier subsystem, which is visualized as the output in figure \ref{fig:sliding_window}. Since the previous layer's output has been converted to classical values post-measurement, the new quantum-classical representation must again be passed through amplitude encoding to map the reduced images to their quantum states. Instead of fragments of the image passed through the reduction block, at this point, the whole image is processed through the amplitude encoding block by a single pass. The input size (the output from the reduction block) is $\frac{N}{r}\times \frac{N}{r}$, and therefore $n = \log_2{((\frac{N}{r})^2)}$ qubits are needed to encode the $(\frac{N}{r})^2$ amplitude values. 
Like the reduction block, the classifier subsystem has a series of quantum ansatzes. It also has extended qubit interaction layers consisting of parameterized rotational gates cascaded with Toffoli gates leveraging 3 qubit interactions. Introducing these gates increases the network's expressibility, resulting in further class separation and improved accuracy. 

After the amplitude encoding block, a composite quantum state of $n$ qubits is created, and the classical data points from $Y_{p,q}$ can now be encoded in the $n$ qubits without further dimension changes. The states are then passed through the first classifier block, a quantum convolutional layer (illustrated as the initial block in figure \ref{fig:decoder_network}). The convolutional layer has the same structure as that of the first convolutional layer in the quantum reduction system illustrated in figure \ref{fig:conv_and_pooling_layers}. The unitary that makes up the convolutional layer ansatz is constructed using the unitary in figure \ref{fig:encoding_conv_ansatz}. The mathematical representation of the unitary is discussed in the previous section and represented mathematically in equation \ref{encoder_ansatz}.

Using the same convolutional unitary as the first block, if each of the patches is of size be $4 \times 4$ $(r=4)$ and $N = 32$, then the resulting reduction network outputs a feature matrix of size $\frac{N}{r} \times \frac{N}{r} = 8 \times 8$. Therefore, the $64$ values can be encoded by amplitude encoding on $n = 6$ qubits of the classifier network. As shown in figure \ref{fig:decoder_network}, the classifier convolutional layer can be expressed using the previous equations in the following way.
\begin{equation}
    U2_{conv} = (U_{1,n} \otimes U_{2,n-1} \otimes ... \otimes U_{\frac{n}{2},\frac{n+2}{2}})\star(U_{1,2} \otimes U_{3,4} \otimes ... \otimes U_{n-1,n})
\end{equation}
where $U_{x,y}$ is the $U$ gate acting on qubits $x$ and $y$, and $n$ is the number of qubits. The quantum state that is processed by $U2_{conv}$ is then made subject to the next block, which is a 3-qubit interaction layer with Toffoli gates and can be denoted by the equation:
\begin{equation}
    U_{c. ansatz-2} = T_{n,n-1, n-2} T_{n-1,n-2, n-3} ... T_{2, n-1, n} T_{1, 2, n}
    \label{c.ansatz2}
\end{equation}
where $T_{x,y,z}$ represents the unitary corresponding to the Toffoli gate with $x$ as the target qubit and $y$, $z$ as the controls. It must be noted the unitaries are adjusted by tensor products with identity matrices $I$ of appropriate dimensions such that they act on the composite quantum state of $n$ qubits. The resultant ansatz, $U_{c. ansantz-2}$ (eqn. \ref{c.ansatz2}), acts on the state output from the previous block $U2_{conv}$ and forwards the state to a series of $R_x$, $R_y$, $R_z$ gates with trainable parameters. $U_{c. ansatz-1}$ comprises these rotational gates and can be expressed by the following relation:
\begin{equation}
    U_{c. ansatz-1} =  (\bigotimes^6 R_z) \star  (\bigotimes^6 R_y) \star (\bigotimes^6 R_x)
\end{equation}
where $\bigotimes^6 R$ represents the 6 rotational gates acting individually on the $n = 6$ qubits. Indeed, marked as c. ansatz 1 in figure \ref{fig:decoder_network}, the $R_x$, $R_y$, and $R_z$ gates are cascaded to develop the final ansatz. The composite state of $n$ qubits is then passed on to the same ansatz block as $U_{c. ansatz-2}$.
The states are then passed through a pooling-type ansatz where the number of qubits is reduced to half. This pooling ansatz (referred to as c. ansatz 3 in figure \ref{fig:decoder_network}) comprises repeated two-qubit gates, the same as the reducer circuit shown in figure \ref{fig:encoding_conv_ansatz}, traces out the bottom qubit, and produces the first qubit state as its output. The two-qubit gates can be mathematically expressed as the following.

\begin{equation}
    P_{1,2} = Cx_1(\theta_1) \star (I \otimes X_2)  \star  Cz_1(\theta_2)
\end{equation}
and the final pooling ansatz is denoted by the equation:
\begin{equation}
    U_{c.ansatz-3} = \bigotimes_{i=1}^{n} P_{2i-1,2i}.
\end{equation}
Here, the $P_{x,y}$ signifies the application of the unitary $U_{pool}$ on x and y qubits. Hence, one qubit from the two in the pooling ansatz is discarded, similar to the pooling operation in the reduction system (equation \ref{pooling_trace}).

Finally, $A$ additional auxiliary qubits (alongside the data qubits modulated by the output from the previous subsystem), where $A$ is the number of classes, are initialized to the $\ket{0}$ states and introduced in the last stage of the classifier. CNOT gates create two-qubit interactions between the data qubits and the auxiliary qubits. For the example of binary classification on $n$ classifier qubits (now reduced to $\frac{n}{2}$ by pooling), this interaction layer can be expressed by the following.
\begin{equation}
    U_{c. ansatz-4.1} = U_{t-1,3,5} U_{c-3,1} U_{c-5,3} U_{c-A1,3} U_{c-A2,5}. 
\end{equation}
Where $U_{t-1,3,5}$ is a Toffoli gate acting on qubit 1 with qubits 3 and 5 as controls, $U_{c-x,y}$ is the CNOT gate acting on qubit $x$ with $y$ as the control. It must be noted the output qubits from $U_{c. ansatz-3}$, are qubits 1, 3, 5, which interact further with $U_{ansatz-4.1}$, ultimately discarding the $n$ data qubits while the $A$ auxiliary qubits remain. The auxiliary qubits are denoted as qubits $A1=$7 and $A2=$8 (for example, with $n = 6$ classifier data qubits, the auxiliary qubits are 7 and 8). If the number of qubits, $n$, is greater than 6, $n>6$, then a sequence of convolutional and pooling will be required just before introducing the auxiliary qubits to reduce the number of qubits to three before applying $U_{c. ansatz-4.1}$.

Finally, as indicated in the latter part of c. ansatz 4 in figure \ref{fig:decoder_network}, these qubits are subject to several rotational gates before measurement. This purely rotational layer can be expressed mathematically as:
\begin{equation}
    U_{ansatz-4.2} = \{\bigotimes^n I\} \otimes \{(\bigotimes^A R_x(\theta_1)) \star (\bigotimes^A R_y(\theta_2)) \star (\bigotimes^A R_z(\theta_3))\}
\end{equation}
where $\bigotimes^nI$ signifies that the data qubits are now untouched by $U_{ansatz-4.2}$ and $\bigotimes^A R$ represents the rotational gates on the two auxiliary qubits.
Finally, the Pauli-Z expectation value of the auxiliary qubits is measured. Again, just like the reduction system, the Pauli-Z operator's final expectation value depends on the auxiliary qubits' final state, as shown in equation \eqref{expectation_value}.

The predicted values are passed through a softmax function, which produces outputs into a range $\text{softmax}(\mathbf{x})_i \in [0, 1]$ for any arbitrary i\textsuperscript{th} class. Thus, the output vector (considering the values from each auxiliary qubit and constructing a vector) now denotes the probabilistic values of the various classes and can be directly compared to the true one-hot-enoded labels of the image in question. The softmax function and the predicted vector, $\overline{\mathbf{y}}$ for $A$ number of auxiliary qubits are shown mathematically below.
\begin{equation}.
    y_i = \text{softmax}(\mathbf{x})_i = \frac{e^{x_i}}{\sum_{j=1}^{A} e^{x_j}}
\end{equation}

\begin{equation}
    \overline{\mathbf{y}} = \begin{bmatrix} y_1 \\ \vdots \\y_i \\ \vdots \\ y_A \end{bmatrix}
\end{equation}
Therefore, the softmax output is now input to the classical cross-entropy cost function, which calculates the cost between the predicted labels and the one-hot-encoded true labels of the images. Consequently, gradient descent is performed, which updates the parameters of the gates in both the quantum reduction and classifier structure. 
It is important to note that the m-qubit circuit used as a reducer is not unique to each of the original image data's $r \times r$ patch values. The same parameters for all the patches are used and optimized by gradient descent to prepare the best possible representation of the input image.

\subsection{Join Optimization}

The gradient descent performed using the classical cross-entropy cost function optimizes the reducer and classifier network. This section summarizes the overall quantum architecture and mathematically expresses the joint optimization process.

The quantum reducer is denoted by $U_r(\theta_r)$ with parameters $\theta_r$ and a quantum classifier $U_c(\theta_c)$ with parameters $\theta_c$. Hence, sequentially, the quantum processes are the following.

\begin{equation}
|\psi_r\rangle = U_r(\theta_r) E(x_0) \ket{\mathbf{0}}
\end{equation}
where  $E(x_0)$ is encoding the classical pixel values $x_0$, $\ket{\mathbf{0}}$ is the initialized state of the input qubits.
The reducer qubits are measured:
\begin{equation}
x_1 = \bra{0} U_r^\dagger(\theta_r) Z U_r(\theta_r) \ket{\mathbf{0}}
\end{equation}
where $Z$ is the Pauli-Z observable and $x_1$ is the expectation value output. This measured classical value is re-encoded into the classifier by the following equation.
\begin{equation}
\ket{\phi(x_1)} = E(x_1) \ket{\mathbf{0}}
\end{equation}
where $E(x_1)$ is the re-encoding function that transfers the classical reducer output to quantum states. Subsequently, the classifier ansatz acts on the prepared states.
\begin{equation}
\ket{\psi_c} = U_c(\theta_c) |\phi(x)\rangle
\end{equation}
Finally, the measurement of the classifier qubits is expressed by the following equation.
\begin{equation}
f(\mathbf{x}; \theta_r, \theta_c) = \bra{\phi(x_1)} U_c^\dagger(\theta_c) Z U_c(\theta_c) \ket{\phi(x_1)}
\end{equation}
The cross-entropy loss function is then defined as:
\begin{equation}
\mathcal{L}(\theta_r, \theta_c) = -\sum_i y_{\text{true},i} \log \left( f(\mathbf{x}_i; \theta_r, \theta_c) \right)
\end{equation}
For the join optimization process, the gradients of the loss function are calculated with respect to both the reducer and classifier parameters, $\theta_1$ and $\theta_2$, respectively.
\begin{equation}
\nabla_{\theta_r} \mathcal{L}(\theta_r, \theta_c) = \frac{\partial \mathcal{L}}{\partial \theta_r}
\end{equation}
\begin{equation}
\nabla_{\theta_c} \mathcal{L}(\theta_r, \theta_c) = \frac{\partial \mathcal{L}}{\partial \theta_c}
\end{equation}
Finally, the parameters are updated by gradient descent:
\begin{equation}
\theta_r \leftarrow \theta_r - \eta \nabla_{\theta_r} \mathcal{L}(\theta_r, \theta_c)
\end{equation}
\begin{equation}
\theta_c \leftarrow \theta_c - \eta \nabla_{\theta_c} \mathcal{L}(\theta_r, \theta_c)
\end{equation}
where $\eta$ is the learning rate. This process is repeated iteratively to jointly optimize the parameters of the quantum reducer and classifier. This procedure of jointly optimizing two shallow quantum networks to process classical data of large size and keeping decoherence to a minimum is a first to the best of the author's knowledge.

\section{Experimental Results}

\subsection{Dataset}
The Fashion MNIST dataset (\cite{xiao2017fashion}) is used to benchmark the proposed network, which contains greyscale images of various clothing items developed as an extension/alternative to the famous MNIST hand-written digit dataset by the Modified National Institute of Standards and Technology. A binary classification task is performed, which involves a combination of classes 0 (shirt/top), 1 (trousers), 2 (pullover), 3 (dress). It generally consists of ten classes of clothing items with specified divisions in training and testing image data, each size $28 \times 28$. Each class has $6000$ train and $1000$ test images.

\subsection{Simulation Environment}
The simulation is performed on the widely used quantum machine learning library, Pennylane (\cite{bergholm2018pennylane}) version 0.28.0 and coded on Python 3.9.8. The classical optimizer used in the network training is the Nesterov Moment Optimizer (\cite{nesterov}) with a momentum of 0.9. A loop is run through the training process where a batch of 60 images is randomly selected and fed into the quantum-classical network. Each image passes through the reduction block, is converted to classical values to receive an attention mask, and is again converted to quantum states and measured in the last stage. The labels are one-hot-encoded, and auxiliary qubits are used to measure the final values to feed them into the cost function. An adaptive learning rate is used where it is kept at 0.01 before the testing accuracy reaches 90\%, after which it is changed to 0.001. Once the batch has been processed sequentially, as stated above, the cost function sums up the losses to perform gradient descent and updates the parameters of both the reducer and the classifier. This training process is run for 5 epochs, and the testing accuracy and loss are evaluated every 20 iterations within each epoch.

\subsection{Results}
The major applications of the proposed architecture involve acting as a data classifier and as a method of intelligent data reduction. Subsequent sections show that the network performs remarkably well on both tasks. Further, the reducer working on patches to produce the reduced image representations can be coupled with any parameterized classifier (quantum/classical). The latter part of this section explores this by evaluating the network performance using different classifier architectures. In this analysis, one of the classifier networks explored is a fully connected linear layer structure often used in classical machine learning literature. The use of a classical neural network classifier cascaded with a quantum network has been simulated previously in other recent works (\cite{chalumuri2021hybrid}, \cite{yang2021decentralizing}); however, the joint optimization of the quantum network, along with any classical network, is a first to the best of the author's knowledge.

Fashion MNIST is a widely accepted and popular dataset to benchmark classification tasks in the quantum NISQ genre of networks (\cite{tak_hur_boss}, \cite{hu2022quantum}, \cite{shen2024classification}) gaining the reputation of portraying a good estimate of the capability of the new architecture in handling image data and producing classification results. Table \ref{tab:table_1_classifier_1} summarizes the results for the binary classification problem involving selected classes using the same version of the proposed architecture. 

Several architectures using end-to-end quantum networks (with no classical neural layers) have been trained on the Fashion MNIST dataset in recent times. These works use a data pre-processing step using significant classical computational resources, recognizing that the quantum networks in this NISQ era face a common difficulty in processing large classical data. The proposed network achieves remarkably high classification accuracy (table \ref{tab:table_1_classifier_1}) without the use of any such classical pre-processing and, additionally, manages to surpass the accuracy reported in \cite{tak_hur_boss} ($94.30\%$) on the binary classification problem (0 versus 1) by a margin of $2.21\%$. At first glance, this margin may seem like a minute improvement over the previous works. However, it must be realized that the reported accuracy in \cite{tak_hur_boss} is derived from using the classical autoencoder with a latent dimension of 8 and input dimension 784, resulting in the training of 17,416 parameters as a data pre-processing step. These parameters are reported to be trained for 10 epochs after the reduced 8-length data is input into the quantum convolutional neural network. 
In contrast, the proposed network in this work avoids such heavy, computationally costly data-preprocessing steps and proposes a comprehensive method for reducing data and feeding it into the NISQ quantum neural network structure while achieving better classification results.


\subsubsection{Variation in Classes}

\begin{table}[]
\centering
\begin{tabular}{c|c|c|c|c|c}
\hline
\hline
Classes evaluated & Class description                                                & \begin{tabular}[c]{@{}c@{}}Encoder \\ parameters\end{tabular} & \begin{tabular}[c]{@{}c@{}}Classifier\\  parameters\end{tabular} & \begin{tabular}[c]{@{}c@{}}Total \\ parameters\end{tabular} & \begin{tabular}[c]{@{}c@{}}Test \\ Accuracy (\%)\end{tabular} \\ \hline
0 vs 1            & \begin{tabular}[c]{@{}c@{}}T-shirt/top vs\\ trouser\end{tabular} & \multirow{8}{*}{8}                                            & \multirow{8}{*}{22}                                              & \multirow{8}{*}{30}                                         & 96.51                                                         \\ \cline{1-2} \cline{6-6} 
1 vs 2            & \begin{tabular}[c]{@{}c@{}}trouser vs\\ pullover\end{tabular}    &                                                               &                                                                  &                                                             & 95.05                                                         \\ \cline{1-2} \cline{6-6} 
1 vs 3            & \begin{tabular}[c]{@{}c@{}}trouser vs\\ dress\end{tabular}       &                                                               &                                                                  &                                                             & 88.75                                                          \\ \cline{1-2} \cline{6-6} 
2 vs 3            & \begin{tabular}[c]{@{}c@{}}pullover vs\\ dress\end{tabular}      &                                                               &                                                                  &                                                             & 92.75                                                         \\ \hline \hline
\end{tabular}
\caption{The table summarizes the results of using the proposed reducer-classifier architecture to perform binary classification on various classes of the fashion MNIST dataset. Most works, \cite{tak_hur_boss}, \cite{hu2022quantum}, \cite{shen2024classification} use the dataset to benchmark their results on quantum learning networks.}
\label{tab:table_1_classifier_1}
\end{table}


The proposed architecture is tested on a combination of classes from the Fashion MNIST dataset, and the results are summarized in table \ref{tab:table_1_classifier_1}. It can be observed that the architecture performs exceptionally well in different combinations. The most common class combinations to be tested for binary classification in QML literature are the classes 0 and 1 (\cite{tak_hur_boss}), although some architectures have also been tested using classes 2 and 3 (\cite{hu2022quantum}). At this stage, it is essential to observe that the number of parameters trained is only $30$, and the circuit depth is lower than conventional QCNN architecture.

The number of layers gradually decreases in a tree-like structure in quantum convolutional neural networks. Still, it is essential to observe that the depth is bound to increase, given that the network is processing classical data of large size. Any encoding method will require increased qubits as the the number of data points increase. (For example, amplitude encoding will need an extra qubit as the classical data size exceeds $2^m$ values, and angle encoding will demand an extra qubit when the data size exceeds $2m$ values, where $m$ is the number of qubits.) The increase in qubits (and hence the unavoidable increase in depth in QCNNs) is undesirable in the NISQ era for two reasons. First, due to various hardware limitations, there is an upper bound to the number of qubits that can be deployed in these networks. Secondly, the increase in depth means that the quantum circuit runs into a higher risk of quantum decoherence. The method proposed in this paper tackles these problems simultaneously by designing a network that uses a small quantum reducer to work on small patches of a larger image and finally reduces its size for the classifier network to tackle. In this way, the network avoids using a large number of qubits at one go; additionally, it uses a reduced feature matrix as an input to the classifier, limiting the number of qubits required and keeping the networks shallow for both the reducer and classifier.

\begin{table}[h]
\centering
\begin{tabular}{ccccccc}
\hline
\hline
Reducer & Classifier & \begin{tabular}[c]{@{}c@{}}reducer \\ parameters\end{tabular} & \begin{tabular}[c]{@{}c@{}}classifier \\ parameters\end{tabular} & \begin{tabular}[c]{@{}c@{}}total \\ parameters\end{tabular} & \begin{tabular}[c]{@{}c@{}}Test \\ Accuracy (\%)\end{tabular} & \begin{tabular}[c]{@{}c@{}}Joint \\ optimization\end{tabular} \\ \hline
proposed  & proposed            & 8 & 22 & 30 & \textbf{96.51} & \checkmark \\ \hline
proposed  & classical FCC   & 8 & 566 & 574 & \textbf{97.35} & \checkmark \\ \hline
naive q. pool  & proposed   & 4 & 22 & 16 & 94.50 & \checkmark \\ \hline
autoencoder  &  proposed   & 132,160 (10 epochs) & 22 & 132,182 & 95.65 & \ding{55} \\ \hline
autoencoder  &  QCNN (\cite{tak_hur_boss})   & 17,416 (10 epochs) & 51 & 17,467 & 94.30 & \ding{55}\\ \hline
\hline
\end{tabular}
\caption{This table summarizes the results obtained using different classifiers and encoder types. It illustrates the capability of various architectures to be jointly optimized with the reducer and investigates the contribution of the proposed reduction and classifier block towards the results produced in Table \ref{tab:table_1_classifier_1}. The classification accuracies shown are for classes 0 vs 1.}
\label{tab:merged_table_classifier_encoder_types}
\end{table}

\subsubsection{Variation in classifier}

\begin{figure}[h]
    \centering
    \includegraphics[scale=0.7]{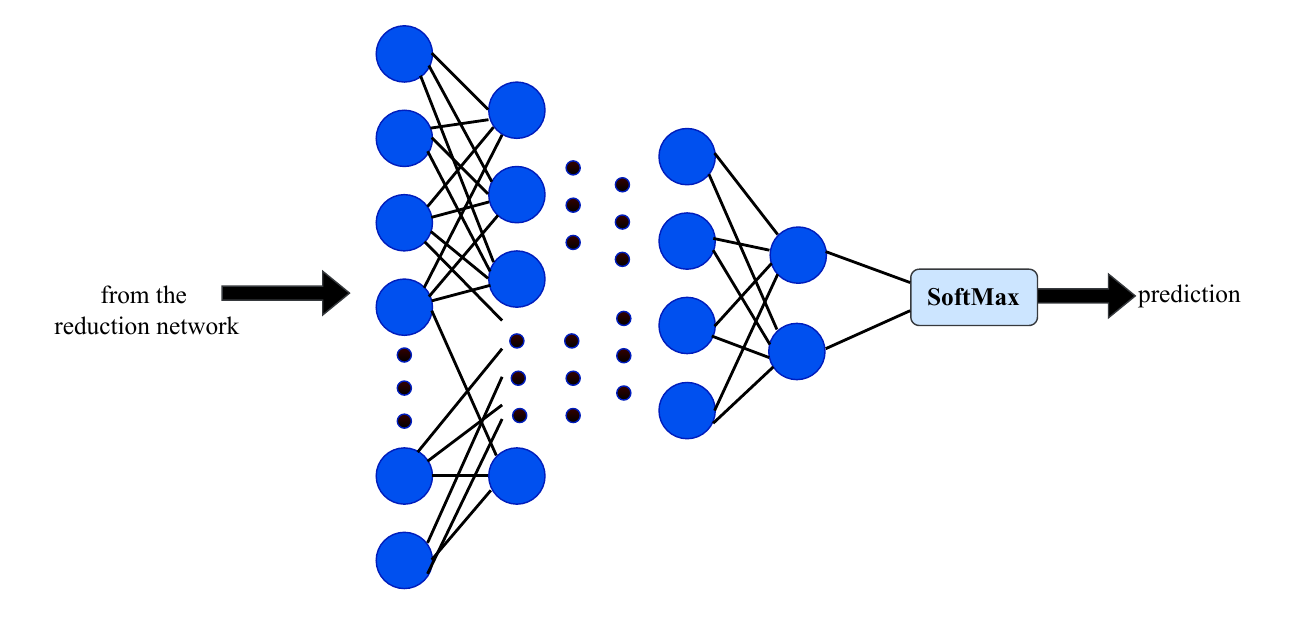}
    \caption{Fully connected classical layer used as a classifier coupled with the proposed reducer. The classifier used in this case had an input size of $64$, which was brought down to $8$, $4$, and finally $2$ in the subsequent fully connected layers. The outputs of the $2$ neurons are then passed through a Softmax function.}
    \label{fig:classical_classifier}
\end{figure}

The general architecture is also tested using a completely different classifier networks to demonstrate its flexibility. The network performs exceptionally well on different classifiers, as shown in table \ref{tab:merged_table_classifier_encoder_types}. Figure \ref{fig:classical_classifier} depicts the general structure of the fully connected (classical neural network) classifier. 

The fully-connected (linear) layer is one of the most straightforward architectures used in the classical neural network literature. It is denoted as the classical FCC (Classifier) in table \ref{tab:merged_table_classifier_encoder_types} and 17,416 training parameters. The structure involves the first layer, which receives the $8 \times 8$ input at $64$ neurons connected to $8$ neurons. The structure progresses with the input size decreasing to 4 and finally to 2, passing through the Softmax block. There may be other combinations of this linear layer that may produce better accuracies; however, the goal of this experiment is not to generate the best possible accuracy but to demonstrate the ability of the proposed reduction network to be optimized with different classifier networks and produce competitive results in all cases. The proposed reducer coupled with the linear layer achieves an exceptionally high test accuracy of 97.35\% with a total trainable parameters of 574, which is still low when compared to classical machine learning architectures.


Using different classifiers and their joint optimization proves that this reduction method can be used in various applications and is not limited to QCNN or other quantum neural networks. The classical fully connected classifier (FCC), shown in table \ref{tab:merged_table_classifier_encoder_types}, is used in \cite{chalumuri2021hybrid}. However, the quantum reducer deployed in the network in \cite{chalumuri2021hybrid} is static and lacks any trainable parameters. The initial network works as a mapper of the classical values into a quantum state, which, upon measurement, assists the deep classical neural network with a multitude of parameters. There is no optimization of quantum gates at any network stage, causing the architecture to function as a quantum-assisted classical machine learning method. The classifier deployed in this work results in a hybrid network like the one proposed in \cite{chalumuri2021hybrid} but with two significant improvements. First and most prominently, the quantum reduction network is trainable and jointly optimized with the classifier. Second, in contrast to the reducer in \cite{chalumuri2021hybrid}, the proposed reduction network includes the self-attention mechanism. Both additions make the proposed architecture much more robust and flexible than the one in \cite{chalumuri2021hybrid}.

\subsubsection{Variation in reducer}


\begin{figure}[h]
    \centering
    \includegraphics[scale=0.7]{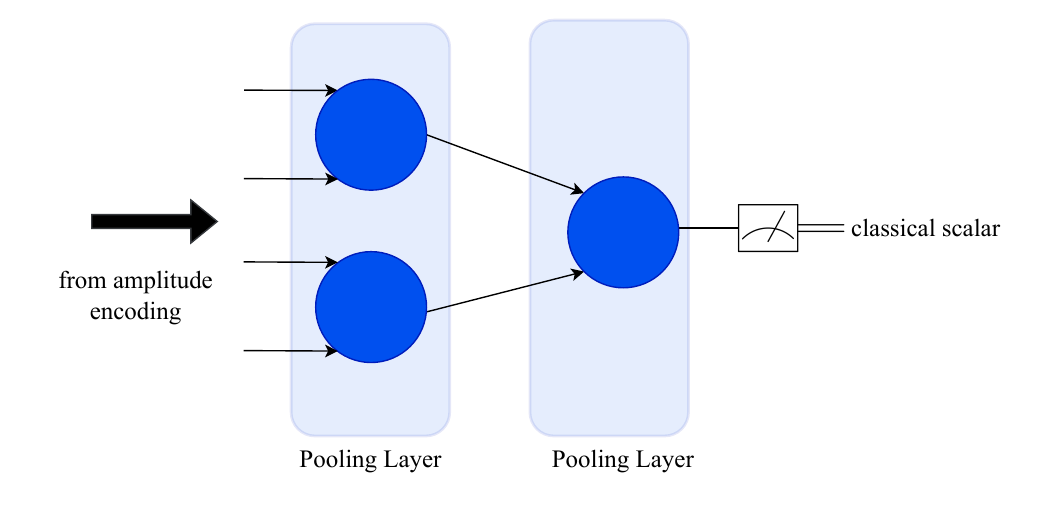}
    \caption{The naive quantum reduction block. This network functions similarly to the proposed reducer, except the reducer network lacks convolutional layers. This means it only relies on a sequence of pooling layers, each with $2$ trainable parameters to reduce the image.}
    \label{fig:q_decoder}
\end{figure}

It has now been shown that the proposed quantum reduction structure coupled with the classifier is exceptionally flexible at learning the dataset and carrying out the classification, achieving excellent classification results. This naturally raises the question regarding the contribution of the reducer and classifier to these results. If the classifier is exceptionally well-designed, the contribution of the reducer may be negligible, implying that many different versions of the quantum reduction block can produce similar results. On the contrary, it might be the reducer whose contribution is more important since it summarizes each image, and the classifier works on its output. A naive quantum reducer that reduces the $r \times r$ pixel values to a single representative value is shown in figure \ref{fig:q_decoder}.  This network does not have the quantum convolutional layers and only depends on the pooling layers with a couple of parameters in each layer to reduce the patch. The architecture constructed using this naive quantum reduction block indicates whether the reducer network significantly contributes to the output, considering the classifier remains unchanged. The findings, as summarized in table \ref{tab:merged_table_classifier_encoder_types}, show that the network achieves a test accuracy of 94.50\% on classes 0 vs. 1. The accuracy is 2.01\% less than the accuracy achieved using the proposed network, which shows that the reduction block has a significant contribution to the network performance. It must be noted that although we are using a naive quantum reducer, the pooling layers are still present and have trainable parameters that are jointly optimized with the classifier network. Therefore, this network structure still enjoys the benefits of joint optimization, which helps it achieve decent accuracy.

Next, a classical autoencoder is trained to reduce the images to a vector of $2^{n}$ so that they can be directly encoded by amplitude encoding in the classifier and entirely bypass the proposed reducer network. It must be noted that the number of parameters to be trained is extensive, with the majority used in the pre-processing step using the autoencoder. The autoencoder has a single latent layer of $64$, precisely the flattened length $(8 \times 8)$ of the reduced representation produced by the quantum reduction block. The autoencoder is trained for $10$ epochs, after which the decoder part is discarded, and the encoder processes the images to produce reduced vectors, which the classifier can then use for training. This method achieves a test accuracy of 95.65\%. The high test accuracy is expected, considering the significant number of parameters involved. However, even after using an autoencoder with 132,160 parameters as a reduction block, the test accuracies are less than the proposed reducer-classifier network, which further displays the ability of the quantum reduction block. Moreover, the autoencoder-classifier structure lacks the joint optimization feature and carries out the reducer and classifier training in two separate training cycles.



\section{Conclusion}
A unique quantum architecture is proposed in this work for processing classical data using NISQ quantum networks for classification. The novel quantum reduction block has been shown to use a patch-based system and an attention-like mechanism to produce reduced representations, which NISQ quantum classifiers can process. Subsequently, join optimization of parameters in both the reducer and classifier is performed, which has shown to bear excellent classification results and solves the imminent problem of processing large classical data using computationally expensive classical mechanisms. Furthermore, the architecture allows the networks to be shallow, effectively arresting quantum decoherence. 


It has been shown that the architecture of the quantum reduction block is robust and can be utilized to process any images of different sizes to be jointly optimized with a multitude of quantum or classical classifier networks. This work has introduced a novel method for processing higher large classical data using small, shallow NISQ quantum networks. When tested on image classification tasks, the proposed method has been shown to supersede computationally costlier classical reduction techniques such as an autoencoder. Moreover, the architecture produces improved classification accuracies compared to recent quantum networks with similar trainable parameters, and the variation in reducers shows that the proposed quantum reduction network is of immense importance in processing images. Therefore, the patch-based processing method coupled with the join optimization significantly improves accuracy numbers.

Future works may include exploring this technique and optimizing it to process different types of extensive data. Moreover, it must be realized that the real benefits of handling quantum decoherence and designing shallow networks that are not demanding in terms of computational resources can be truly realized when running the proposed architecture on a real quantum computer. Finally, the network can be optimized in the future for various applications where NISQ quantum machine learning can make great strides but is only restricted from doing so due to the problem of handling large clasical data.

\subsection{Data availability}

\noindent The Fashion MNIST dataset (\cite{xiao2017fashion}) used in this paper is publicly available in \href{https://github.com/zalandoresearch/fashion-mnist?tab=readme-ov-file}{this repository}.

\subsection{Code availability}

\noindent The source code for the proposed architecture will be made available after publication.

\subsection*{Acknowledgments}\noindent The authors thank Sowmitra Das for his insightful discussions and advice. The gratitude extends to Raisa Mashtura and MD. Jahin Alam for their valuable suggestions regarding this work.

\subsection*{Author Contributions}\noindent J.M. and S.A.F developed the architecture of the network. J.M. wrote the code for the proposed architecture and generated the results. J.M. and S.A.F. wrote the manuscript.

\subsection*{Competing interests}\noindent The authors declare no competing interests.




\bibliography{bibliography}

\begin{thebibliography}{10}

\bibitem{cong_quantum_org}
Iris Cong, Soonwon Choi, and Mikhail~D Lukin.
\newblock Quantum convolutional neural networks.
\newblock {\em Nature Physics}, 15(12):1273--1278, 2019.

\bibitem{Bausch2020RecurrentQN}
Johannes Bausch.
\newblock Recurrent quantum neural networks.
\newblock {\em ArXiv}, abs/2006.14619, 2020.

\bibitem{cherrat2022quantum}
El~Amine Cherrat, Iordanis Kerenidis, Natansh Mathur, Jonas Landman, Martin Strahm, and Yun~Yvonna Li.
\newblock Quantum vision transformers.
\newblock {\em arXiv preprint arXiv:2209.08167}, 2022.

\bibitem{yanofsky2008quantum}
N.S. Yanofsky and M.A. Mannucci.
\newblock {\em Quantum Computing for Computer Scientists}.
\newblock Cambridge University Press, 2008.

\bibitem{farhi2018classification}
Edward Farhi and Hartmut Neven.
\newblock Classification with quantum neural networks on near term processors.
\newblock {\em arXiv preprint arXiv:1802.06002}, 2018.

\bibitem{schuld2020circuit}
Maria Schuld, Alex Bocharov, Krysta~M Svore, and Nathan Wiebe.
\newblock Circuit-centric quantum classifiers.
\newblock {\em Physical Review A}, 101(3):032308, 2020.

\bibitem{liu2018differentiable}
Jin-Guo Liu and Lei Wang.
\newblock Differentiable learning of quantum circuit born machines.
\newblock {\em Physical Review A}, 98(6):062324, 2018.

\bibitem{perez2020data}
Adri{\'a}n P{\'e}rez-Salinas, Alba Cervera-Lierta, Elies Gil-Fuster, and Jos{\'e}~I Latorre.
\newblock Data re-uploading for a universal quantum classifier.
\newblock {\em Quantum}, 4:226, 2020.

\bibitem{maccormack2022branching}
Ian MacCormack, Conor Delaney, Alexey Galda, Nidhi Aggarwal, and Prineha Narang.
\newblock Branching quantum convolutional neural networks.
\newblock {\em Physical Review Research}, 4(1):013117, 2022.

\bibitem{opt-2qubit}
Farrokh Vatan and Colin Williams.
\newblock Optimal quantum circuits for general two-qubit gates.
\newblock {\em Physical Review A}, 69(3):032315, 2004.

\bibitem{sim2019expressibility}
Sukin Sim, Peter~D Johnson, and Al{\'a}n Aspuru-Guzik.
\newblock Expressibility and entangling capability of parameterized quantum circuits for hybrid quantum-classical algorithms.
\newblock {\em Advanced Quantum Technologies}, 2(12):1900070, 2019.

\bibitem{lung_cancer}
Siddhant Jain, Jalal Ziauddin, Paul Leonchyk, Shashibushan Yenkanchi, and Joseph Geraci.
\newblock Quantum and classical machine learning for the classification of non-small-cell lung cancer patients.
\newblock {\em Springer Nature Applied Sciences}, 2(6):1--10, 2020.

\bibitem{weather}
Graham~R Enos, Matthew~J Reagor, Maxwell~P Henderson, Christina Young, Kyle Horton, Mandy Birch, and Chad Rigetti.
\newblock Synthetic weather radar using hybrid quantum-classical machine learning.
\newblock {\em arXiv preprint arXiv:2111.15605}, 2021.

\bibitem{chemical}
O~Anatole Von~Lilienfeld.
\newblock Quantum machine learning in chemical compound space.
\newblock {\em Angewandte Chemie International Edition}, 57(16):4164--4169, 2018.

\bibitem{yang2022semiconductor}
Yuan-Fu Yang and Min Sun.
\newblock Semiconductor defect detection by hybrid classical-quantum deep learning.
\newblock In {\em Proceedings of the IEEE/CVF Conference on Computer Vision and Pattern Recognition}, pages 2323--2332, 2022.

\bibitem{yang2021decentralizing}
Chao-Han~Huck Yang, Jun Qi, Samuel Yen-Chi Chen, Pin-Yu Chen, Sabato~Marco Siniscalchi, Xiaoli Ma, and Chin-Hui Lee.
\newblock Decentralizing feature extraction with quantum convolutional neural network for automatic speech recognition.
\newblock In {\em ICASSP 2021-2021 IEEE International Conference on Acoustics, Speech and Signal Processing (ICASSP)}, pages 6523--6527. IEEE, 2021.

\bibitem{ballard1987modular}
Dana~H Ballard.
\newblock Modular learning in neural networks.
\newblock In {\em Aaai}, volume 647, pages 279--284, 1987.

\bibitem{abdi2010principal}
Herv{\'e} Abdi and Lynne~J Williams.
\newblock Principal component analysis.
\newblock {\em Wiley interdisciplinary reviews: computational statistics}, 2(4):433--459, 2010.

\bibitem{tak_hur_boss}
Tak Hur, Leeseok Kim, and Daniel~K Park.
\newblock Quantum convolutional neural network for classical data classification.
\newblock {\em Quantum Machine Intelligence}, 4(1):1--18, 2022.

\bibitem{fan2023hybrid}
Fan Fan, Yilei Shi, Tobias Guggemos, and Xiao~Xiang Zhu.
\newblock Hybrid quantum-classical convolutional neural network model for image classification.
\newblock {\em IEEE transactions on neural networks and learning systems}, 2023.

\bibitem{superdense}
Charles~H. Bennett and Stephen~J. Wiesner.
\newblock Communication via one- and two-particle operators on einstein-podolsky-rosen states.
\newblock {\em Phys. Rev. Lett.}, 69:2881--2884, Nov 1992.

\bibitem{schuld2018supervised}
Maria Schuld and Francesco Petruccione.
\newblock {\em Supervised learning with quantum computers}, volume~17.
\newblock Springer, 2018.

\bibitem{grant2018hierarchical}
Edward Grant, Marcello Benedetti, Shuxiang Cao, Andrew Hallam, Joshua Lockhart, Vid Stojevic, Andrew~G Green, and Simone Severini.
\newblock Hierarchical quantum classifiers.
\newblock {\em npj Quantum Information}, 4(1):65, 2018.

\bibitem{xiao2017fashion}
Han Xiao, Kashif Rasul, and Roland Vollgraf.
\newblock Fashion-mnist: a novel image dataset for benchmarking machine learning algorithms.
\newblock {\em arXiv preprint arXiv:1708.07747}, 2017.

\bibitem{bergholm2018pennylane}
Ville Bergholm, Josh Izaac, Maria Schuld, Christian Gogolin, Shahnawaz Ahmed, Vishnu Ajith, M~Sohaib Alam, Guillermo Alonso-Linaje, B~AkashNarayanan, Ali Asadi, et~al.
\newblock Pennylane: Automatic differentiation of hybrid quantum-classical computations.
\newblock {\em arXiv preprint arXiv:1811.04968}, 2018.

\bibitem{nesterov}
Yu~E Nesterov.
\newblock A method for solving the convex programming problem with convergence rate.
\newblock In {\em Dokl. Akad. Nauk SSSR,}, volume 269, pages 543--547, 1983.

\bibitem{chalumuri2021hybrid}
Avinash Chalumuri, Raghavendra Kune, and BS~Manoj.
\newblock A hybrid classical-quantum approach for multi-class classification.
\newblock {\em Quantum Information Processing}, 20(3):119, 2021.

\bibitem{hu2022quantum}
Zhirui Hu, Peiyan Dong, Zhepeng Wang, Youzuo Lin, Yanzhi Wang, and Weiwen Jiang.
\newblock Quantum neural network compression.
\newblock In {\em Proceedings of the 41st IEEE/ACM International Conference on Computer-Aided Design}, pages 1--9, 2022.

\bibitem{shen2024classification}
Kevin Shen, Bernhard Jobst, Elvira Shishenina, and Frank Pollmann.
\newblock Classification of the fashion-mnist dataset on a quantum computer.
\newblock {\em arXiv preprint arXiv:2403.02405}, 2024.

\end{thebibliography}

\bibliographystyle{unsrt}

\end{document}